\documentstyle{article}
\newcommand{\beq}{\begin{equation}}
\newcommand{\eeq}{\end{equation}}
\newcommand{\beqa}{\begin{eqnarray}}
\newcommand{\eeqa}{\end{eqnarray}}
\newcommand{\nn}{\nonumber}

\newcommand{\half}{\frac{1}{2}}

\newcommand{\xt}{\tilde{X}}

\newcommand{\uind}[2]{^{#1_1 \, ... \, #1_{#2}} }
\newcommand{\lind}[2]{_{#1_1 \, ... \, #1_{#2}} }

\newcommand{\pbr}[2]{ \{ \hspace*{-2.2pt} [ #1 , #2 ] \hspace*{-2.5pt} \} }
\newcommand{\we}{\wedge}
\newcommand{\dv}{d^V}
\newcommand{\nbrpq}[2]{\nbr{\xxi{#1}{1}}{\xxi{#2}{2}}}

\newcommand{\rbox}[2]{\raisebox{#1}{#2}}
\newcommand{\xx}[1]{\raisebox{1pt}{$\stackrel{#1}{X}$}}
\newcommand{\xxi}[2]{\raisebox{1pt}{$\stackrel{#1}{X}$$_{#2}$}}

\newcommand{\ff}[1]{\raisebox{1pt}{$\stackrel{#1}{F}$}}
\newcommand{\dd}[1]{\raisebox{1pt}{$\stackrel{#1}{D}$}}
\newcommand{\nbr}[2]{{\bf[}#1 , #2{\bf ]}}
\newcommand{\der}{\partial}

\newcommand{\Om}{\Omega}
\newcommand{\om}{\omega}
\newcommand{\eps}{\epsilon}

\newcommand{\inn}{\hspace*{2pt}\raisebox{-1pt}{\rule{6pt}{.3pt}\hspace*
{0pt}\rule{.3pt}{8pt}\hspace*{2pt}}}

\newcommand{\bm}{\boldmath}
\newcommand{\vol}{\widetilde{vol}}

\newcommand{\bd}{\mbox{\bm $d$}}

\newcommand{\bI}{\mbox{\bm $I$}}

\title{Basic structures of the covariant canonical formalism for
fields based on the De Donder--Weyl theory\thanks{Invited
contribution to be published
in the volume devoted to the 90-th jubilee of
Prof. D. D. Ivanenko, ed. G. Sardanashvily,
Moscow University Publ., Moscow, 1994} }

\author{ Igor V. Kanatchikov\thanks{\it e-mail:
igor@acds16.physik.rwth-aachen.de}
\\ \it Institut f\"ur
Theoretische Physik E\\
\it RWTH Aachen \\ \it D-52056 Aachen, Germany }
\date{
October 20, 1994}

\begin{document}

\maketitle

\vspace*{-80mm}

\begin{flushright}
PITHA 94/47 \\
October 1994\\
hep-th/9410238\\
\end{flushright}

\vspace*{60mm}

\begin{abstract}
We discuss a field theoretical extension of the basic
structures of classical analytical mechanics within the
framework of the De Donder--Weyl (DW) covariant Hamiltonian
formulation. The analogue of the symplectic form is argued
to be the {\em polysymplectic} form of degree $(n+1)$,
where
$n$ is the dimension of space-time, which defines a map
between multivector fields or, more generally,
graded derivation operators on exterior algebra,
and forms of various degrees
which play
a role of dynamical variables.
The Schouten-Nijenhuis bracket on multivector fields
induces the graded analogue of the Poisson bracket on forms,
which turns the exterior algebra of (horizontal)
forms to a Gerstenhaber algebra. The equations of motion
are written in terms of the Poisson bracket on forms
and it is argued that the bracket with $H\vol$,
where $H$ is the DW Hamiltonian function
and $\vol$ is the horizontal
(i.e. space-time) volume form, is related to the
operation of exterior differentiation of forms.
\end{abstract}

\begin{quote}

\end{quote}

\section{Introduction}

It has long been known that the Euler-Lagrange field
equations may be written in the following covariant
form reminiscent the Hamilton's form of equations of
motion in mechanics:
\beq
\frac{\der p^i_a}{\der x^i}=-\frac{\der H}{\der y^a},
\hspace*{50pt}
\frac{\der y^a}{\der x^i}=\frac{\der H}{\der p^i_a} .
\eeq
Here $\{x^i\}$ $(i=1, ... , n)$ are space-time coordinates,
$\{ y^a \}$ $(a=1,...,m)$ are field variables and the
quantities $p^i_a$ and $H$ are given by the Lagrangean
density $L=L(y^a,\der_i y^a, x^i)$ as follows:
\beqa
p^i_a &:=& \frac{\der L}{\der(\der_i y^a)} , \\
H&:=&p^i_a\der_iy^a -L .
\eeqa
This formulation
is known form the approaches of De Donder,
Weyl  and some others (see e.g.
\cite{Rund, Kastrup83} for a review and further references)
to the variational calculus of multiple integrals. This,
together with a close similarity to the Hamilton's equations
in mechanics, is
the reason why we call the formulation above the
{\em De Donder--Weyl (DW) Hamiltonian formulation\/}.
We shall also call $n$ quantities $p^i_a$ associated
to each field variable $y^a$ the {\em DW canonical momenta\/},
and the scalar density $H$ is to be refered to as the
{\em DW Hamiltonian function\/}.

The aforementioned Hamiltonian formulation  of field equations
has two major advantages over the standard Hamiltonian
formalism in field theory -- manifest covariance, in the
sense that space and time variables enter the formulation on
a completely equal footing, and  finite dimensionality, in the
sense that the analogue of the phase space, the space of
variables $(y^a,p^i_a,x^i)$ which we call the
{\em DW phase space\/}, is finite dimensional.
Because of
the latter circumstances we may also call the
Hamiltonian formulation above
covariant and finite dimensional.

It is well known that the standard  Hamiltonian formulation of
equations of motion, both in mechanics and field theory,
reveals the structures, like the Poisson brackets or
the symplectic structure, which are important for
transition to a quantum description of the dynamics
along the lines of canonical or geometric quantization.
The analogous structures in the DW Hamiltonian formulation
are, essentially, not known, as well as the answer to the
question whether it is possible to
construct a quantum field
theoretical formalism based on the finite dimensional
canonical framework in field theory.

The latter problem was originally discussed
by Born \cite{Born34} and Weyl \cite{Weyl34} in 1934.
Then, certain interest to it arose again in the
beginning of seventies, when a substantial
progress was made in understanding the differential
geometric structures behind the DW canonical theory
for fields (see e.g. [5-11]),
which culminated in the so-called
multisymplectic formalism principally formulated in
\cite{Kij ea} (see [12-16] for recent developments).
In this connection
interesting ideas were suggested then by Guenther [17]
and, more recently, Sardanashvily discussed possible
applications of his
closely related "multimomentum Hamiltonian
formalism" to quantum field theory \cite{Sardan}.

In this paper we concisely review our recent
study of the canonical
structure of the De Donder--Weyl Hamiltonian
formulation in field theory \cite{Kan, Kan in prep}.
The aim of this research was to reveal those
structures within the covariant finite dimensional
canonical formalism for fields
which generalize or are analogous
to the structures in the formalism of classical mechanics
which are known to be important for  quantization.
Compared to \cite{Kan},
Sect. 7 containes new results which will be
presented in more details  elsewhere.

\section{A heuristic consideration}

A construction of the canonical scheme corresponding
to the DW Hamiltonian formulation faces a difficulty
related to the fact that the number of field
variables and the associated DW momenta is not equal,
whereas the equality of the number of generalized
coordinates and the conjugate momenta is an
unavoidable prerequisite for the symplectic or
the Poisson structure. Another problem is the
suitable identification of the "evolution operation"
generalizing the total time derivative in the
canonical form of equations of motion in mechanics.
In fact, in the left-hand side of the first of the DW
Hamiltonian equations, eqs. (1),
one has a space-time gradient while in the
l.h.s of the second one there stands a space-time divergence.
How to unify these two operations?

One can overcome these
difficulties by noticing  that the DW Hamiltonian equations
may be written in terms of the form of degree $(n-1)$
associated with the DW momenta
\[ p_a := p_a^i \der_i \inn \vol \]
in the following "unified" form
\beqa
dy^a&=&"\frac{\der (H\vol)}{\der p_a}":=\frac{\der H}{\der p^i_a} dx^i  , \nn
\\
dp_a&=&-\frac{\der (H\vol)}{\der y^a}  , \nn
\eeqa
where $\vol:=dx^1\we ... \we dx^n$ is the volume-form on the
space-time manifold (for simplicity we use the coordinate systems
with the unit metric determinant) and the expression in the
quotation marks  involves a formal differentiation
of the $n$-form
with respect to the $(n-1)$-form, which may be
attributed certain sense with the help of
the identity
$\vol\ \delta^i_j = dx^i \we \der_j \inn \vol$.
The formulation above suggests that the exterior differential may
be a suitable generalization of the time derivative, and that the
pair of variables $(y^a, p_a)$,  one of which is a $0$-form
and another one is a $(n-1)$-form, may be a suitable analogue
of the canonically conjugate variables. The latter feature
means that the Poisson bracket yet to be defined has to
act on forms of various degrees
and
that the analogue of the canonical transformations may
mix the forms of different degrees. All these properties are
revealed in the formalism we present below.

\section{Poincar\'e-Cartan form, classical extremals and
the polysymplectic form}

As a starting point of our construction we take the
field theoretical analogue of the Poincar\'e--Cartan form which,
being the fundamental object of the geometric approach to the
calculus of variation, contains all the necessary information
for constructing a canonical scheme.
It is known that all the elements of the canonical formulation
of mechanics may be recovered step by step from the
Poincar\'e--Cartan
form (see e.g. \cite{Arnold}). We present here a
field theoretical
generalization of these procedures.

The multidimensional or field theoretical analogue of the
Poincar\'e--Cartan form written in terms of the DW Hamiltonian
variables is the $n$--form  (see e.g. [2,8,9,12-16,22] )
\beq
\Theta=p^i_a dy^a\we \om_i - H\omega ,
\eeq
where $\widetilde{vol}:=
dx^1\wedge...\wedge dx^n=:\omega $ and $\om_i:=\der_i \inn \vol$.
The exterior differential of the Poincar\'e--Cartan form,
the canonical $n+1$--form
\beq
\Om_{DW}:=d\Theta ,
\eeq
is of\ fundamental importance since the classical extremals
are known to be its isotropic $n$--hypersurfaces in the
$(m+mn+n)$--dimensional
DW phase space with the coordinates
$z^M:=(y^a, p^i_a, x^i)$
$(M=1,...,m,m+1,...,m+mn,...,m+mn+n)$.
If we describe these hypersurfaces as the
integral surfaces of the multivector field of degree
$n$, $\xx{n}$,
\beq
\mbox{$\xx{n}:=\frac{1}{n!} X$$^{M_1...M_n}(z)\,\der
\lind{M}{n} , $}
\eeq
where $\der \lind{M}{n} := \partial_{M_1}
\wedge...\wedge\partial_{M_n},$
which are defined as the solutions $z^M=z^M(x)$ of
the equations
\begin{equation}
\mbox{$\stackrel{n}{X}$$^{M_1...M_n}$}(z)=
{\cal N}
\frac{\partial(z^{M_1},...,z^{M_n})}
{\partial(x^1,...,x^n)},
\end{equation}
with ${\cal N}$ depending on the parametrization of the
surface,
then the condition on $\xx{n}$ to give the extremals is that it
annihilates the canonical $(n+1)$ form, that is
\beq
\xx{n} \inn\ \Omega_{DW}=0.
\eeq
This condition specifies a part of the components of $\xx{n}$
and using the parametrization (7)
of its integral surfaces one arrives at
the DW Hamiltonian equations, eqs. (1).
One assumes that the DW extended phase space is a
globally trivial bundle over the space-time (=horizontal
subspace) with a typical fibre called
a vertical subspace or a DW phase space
with coordinates $z^v:=(y^a,p_a^i)$
In these terms,
eq. (8) specifies, in fact, only the vertical,
$X^v{}\uind{i}{n-1}$, and a part of bivertical,
$X^a{}^i_b{}\uind{i}{n-2}$,
components of $\xx{n}$.
However, the equation following from the bivertical
components may be shown to be a consequence of the
DW Hamiltonian equations following from the vertical
components.
Thus, the vertical components of the canonical multivector
field $\xx{n}$ are sufficient for describing
the classical dynamics.

Therefore, we introduce the notion of
{\em vertical\/} multivector of
degree $p$, $\xx{p}{}^V$,
having a component form
\beq
\xx{p}{}^V:=  \frac{1}{(p-1)!} X^v{}\uind{i}{p-1}\der_v{}\lind{i}{p-1}  .
\eeq
Then from (8) it follows that the DW canonical field
equations may be recovered from
the following condition on the vertical $n$-vector field:
\beq
\xx{n}{}^V \inn\ \Omega = (-)^n d^V H,
\eeq
provided the parametrization in (7) is choosen in such a way that
$\frac{1}{n!}\xx{n}{}\uind{i}{n}\der\lind{i}{n}\inn\ \om = 1$.
Here $\dv$
is the vertical exterior differential:
$\dv ... := dz^v\we\der_v \, ...  $
and the form
\beq
\Omega := -dy^a\we dp^i_a \we \om_i,
\eeq
which is defined as the vertical exterior differential of the
vertical (=non-horizontal) part of the
Poincar\'e-Cartan form:
$
\Omega :=\dv \Theta^V, 
\Theta^V:=p^i_a dy^a\we \om_i ,
$
is to be refered to as the
{\em polysymplectic\/} form.

\section{Hamiltonian multivector fields and forms}

Eq. (10) may be viewed as a map from $0$--forms to vertical $n$--vectors.
In general, the polysymplectic form maps horizontal
$p$--forms $\ff{p}$ $(p=0,...,n-1)$, which are defined to
have a form
\beq
\ff{p}:=\frac{1}{p!}F_{i_1 ... i_p}(z) dx\uind{i}{p},
\eeq
where $ dx\uind{i}{p}:=dx^{i_1}\wedge ...\wedge dx^{i_p}$,
to vertical multivectors of degree $q=n-p$, $\xx{q}_F$:
\beq
\xx{q}_F \inn \Omega = \dv \ff{p}.
\eeq
We call {\em Hamiltonian\/} the vertical multivector field
fulfilling (13) for some horizontal form $F$.
Similarly, the horizontal forms to which a vertical multivector
can be associated through the map (13) are to be called the
{\em Hamiltonian forms}. The multivector field $\xx{p}_F$ is
refered to as the Hamiltonian $p$--(multi)vector field associated
with the form $F$.

Note that the map (13) given by the polysymplectic form has a
non-trivial kernel formed by the Hamiltonian multivector fields
$\xx{p}_0$ fulfilling
\beq
\xx{p}_0 \inn \Omega = 0,
\eeq
which are to be refered to as the
{\em primitive\/} Hamiltonian fields.
It is apparent that  the Hamiltonian multivector field associated
with a given form is determined
up to
an arbitrariness related to
addition of a primitive field of the same degree. Therefore,
the map (13) should be rather viewed as a map of forms
to the equivalence classes
of Hamiltonian multivector fields modulo addition of primitive fields.
The quotient space of Hamiltonian multivector fields modulo primitive
ones is the space on which the polysymplectic form may be considered
as nondegenerate.

On the other hand,
the very existence of the map from forms to
multivector fields imposes restrictions
on the forms themselves. These restrictions are implicit in the notion
of Hamiltonian form. Let us consider an example of a form of degree
$(n-1)$, $F:=F^i\der_i \inn\ \vol$ to which a Hamiltonian vector field
$X_F:=X^a\der_a + X^i_a \der^a_i$ is associated by the map
$X_F\inn \Omega = \dv F$ or, in terms of components,
$X^v\der_v \inn (-dy^a\we dp^i_a\we \om_i)=
(\der_aF^idy^a+\der^a_jF^idp^j_a)\we \om_i ,  $
whence it follows
\beqa
X^i_a&=&\der_aF^i , \\
-X^a\delta^i_j&=&\der^a_j F^i  .
\eeqa
It is clear that the latter equation imposes a restriction
on an
admissible dependence of the components of $F$ on the DW momenta.
Using the integrability condition of (16)
one discloses that
the most general admissible form of $F^i$ compatible with (16)
is
\beq
F^i(y,p,x)= f^a(y,x)p^i_a\om_i + g^i(y,x) .
\eeq

\section{Generalized canonical symmetry and the brackets of
Hamiltonian multivector fields and forms}

The hierarchy of maps (13) may be viewed as a local consequence
(which hold globally for Hamiltonian multivector fields)
of the following hierarchy of symmetries of the polysymplectic form
\beq
\mbox{$\pounds$\rbox{1pt}{$_{\stackrel{p}{X}}$} $\Omega = 0 $}
\quad (p=1,...,n)
\eeq
which are expressed in terms of the
generalized Lie derivatives with
respect to the vertical multivector fields
\beq
\mbox{$\pounds$\rbox{1pt}{$_{\stackrel{p}{X}}$} $\mu :=$
\rbox{1pt}{$\stackrel{p}{X}$}\inn\
$d^V\mu -(-1)^p \, d^V$(\rbox{1pt}{$\stackrel{p}{X}$}\inn\ $\mu)$}.
\eeq
The graded symmetry above extends in an apparent way
the canonical symmetry of the
symplectic form known from mechanics.
We call the multivector fields fulfilling (18) {\em locally
Hamiltonian\/}.

The notion of Lie derivative with respect to a multivector
field gives rise to the bracket operation on locally
Hamiltonian multivector fields:
\beq
\mbox{\nbrpq{p}{q}} \inn\ \Omega :=
\mbox{$\pounds$\rbox{1pt}{$_{\stackrel{p}{X}{}_1}$}}
(\xx{q}{}_2 \inn\ \Omega) .
\eeq
It is easily revealed that
\beqa
\mbox{$deg($\nbrpq{p}{q})} &=& p+q-1,    \\
\mbox{\nbrpq{p}{q}} &=& -(-1)^{(p-1)(q-1)}
\nbr{\xxi{q}{2}}{\xxi{p}{1}} ,   \\
\mbox{$(-1)^{g_1 g_3}$\nbr{\xx{p}}{\nbr{\xx{q}}{\xx{r}}}}
&+& 
\mbox{$(-1)^{g_1 g_2}$\nbr{\xx{q}}{\nbr{\xx{r}}{\xx{p}}}} \nn \\
&+&\mbox{$(-1)^{g_2 g_3}$ \nbr{\xx{r}}{\nbr{\xx{p}}{\xx{q}}} $=0,$}
\eeqa
where $g_1=p-1, \; g_2=q-1$ and $g_3=r-1$.
Therefore, the bracket above
is the
Schouten-Nijenhuis (SN) bracket of vertical multivector fields
\cite{SN}. Clearly, the set of locally
Hamiltonian multivector fields is a graded Lie algebra with
respect to the SN bracket.

The SN bracket of two Hamiltonian multivector fields
gives rise to the Poisson bracket of the Hamiltonian
forms they are associated with:
\begin{eqnarray}
\mbox{\nbrpq{p}{q}\inn\ $\Omega$} & =\, &
\mbox{\mbox{$\pounds$\rbox{1pt}{$_{\stackrel{p}{X}{}_1}$}}
$d^{V}$\ff{s}$_{2}$} \nonumber \\
                                  & =\, &
\mbox{$ (-1)^{p+1}\, d^{V}($\xxi{p}{1}\inn\ $d^{V}$\ff{s}$_{2})$} \\
                                  & =:  &
\mbox{$- d^{V} \pbr{\ff{r}_{1}}{\ff{s}_{2}},$} \nonumber
\end{eqnarray}
where $r=n-p$ and $s=n-q$. From the definition of the
bracket of Hamiltonian forms it follows
\beq
\mbox{$\pbr{\ff{r}_1}{\ff{s}_2} = (-1)^{(n-r)}X_{1} \inn\ d^{V} \ff{s}_2
= (-1)^{(n-r)}X_{1} \inn\ X_{2} \inn\  \Omega.$    }
\eeq
One can easily prove the following properties of the Poisson
bracket defined above
\begin{quote}
(i) degree
\beq
\mbox{$deg \pbr{\ff{r}_1}{\ff{s}_2} = r+s-n+1$,}
\eeq
(ii) graded antisymmetry
\beq
\mbox{$\pbr{\ff{r}_1}{\ff{s}_2} = -(-1)^{\sigma}
\pbr{\ff{s}_2}{\ff{r}_1},$ }
\eeq
where $\sigma=(n-r-1)(n-s-1)$.\\
(iii) graded Leibniz rule
\beq
\mbox{$\pbr{\ff{p}}{\ff{q} \wedge \ff{r}} = \pbr{\ff{p}}{\ff{q}} \wedge
\ff{r} + (-1)^{q(n-p-1)} \ff{q} \wedge \pbr{\ff{p}}{\ff{r}}, $   }
\eeq
(iv) graded Jacobi identity
\end{quote}
\begin{eqnarray}
\mbox{$(-1)^{g_1 g_3} \pbr{\ff{p}}{\pbr{\ff{q}}{\ff{r}}}$}  &+& \nonumber \\
\mbox{$(-1)^{g_1 g_2} \pbr{\ff{q}}{\pbr{\ff{r}}{\ff{p}}}$} &+&
\mbox{$(-1)^{g_2 g_3} \pbr{\ff{r}}{\pbr{\ff{p}}{\ff{q}}} $}= 0,
\end{eqnarray}
where $g_1 = n-p-1$, $g_2 = n-q-1$ and $g_3 = n-r-1$.
Therefore,
the Poisson bracket defined in (24)
equippes
the set of Hamiltonian forms
with the structure of a Gerstenhaber algebra
\cite{Gerstenhaber}.

\section{The bracket form of equations of motion}

The analogy with the Hamiltonian formalism in mechanics
suggests that the bracket with the DW Hamiltonian
function $H$ is related to the equation of motion.
For the bracket of a
$(n-1)$-form $F:=F^i\om_i$ with $H$ one has
\[\pbr{H}{F}=X_F\inn\ \dv H=X_F{}^a\der_a H+
X_F{}^i_a \der^a_iH \]
where the components of $X_F$ are given by (15) and (16).
Using DW Hamiltonian equations, eqs. (1), and
introducing the notion of the {\em total} exterior differential
$\bd$ of a horizontal  form of degree $p$, $\ff{p}$:
\[\bd \ff{p} := \der_iz^M dx^i\we\der_M\ff{p}
=\der_iz^vdx^i\we\der_v\ff{p}+dx^i\we \der_i\ff{p}
= \bd^V F + d^{hor} F \]
one arrives at the following bracket representation
of the equation of motion of a Hamiltonian form of
degree $(n-1)$
\beq
*^{-1} \bd F=\pbr{H}{F}+\der_iF^i,
\eeq
where one makes use of the
inverse Hodge duality operation: $*^{-1}\om := 1$.
The last term accounts for a possible explicit
dependence of $F^i$ on the space-time coordinates.

\medskip

The generalization  of the previous result to forms of
arbitrary degree requires a certain extension of the
canonical scheme developed above
in sect. (4) and (5).
In fact, because the bracket of any $\ff{p}$ with $H$
vanishes identically when $p<n-1$ the only way out is to
attribute certain meaning to the Poisson bracket with
the DW Hamiltonian $n$-form $H\om$.
Therefore, we extend the hierarchy of maps (13)
by including the $n$-forms
\beq
\xt_F\inn\ \Omega = \dv \ff{n}.
\eeq
The formal degree of the object $\xt_F$ associated with
an $n$-form is zero and one identifies it with the
vertical vector-valued horizontal one-form $\xt$,
$\xt := \xt^v{}_i\, dx^i \otimes \der_v$,
acting on $\Omega$ through the Fr\"olicher-Nijenhuis inner
product \cite{FN} (see also \cite{Kolar})
which is defined as follows
\beq
\mbox{$\tilde{X}\inn\ \Omega := X^{v}_{\cdot\, k} dx^{k}
\wedge\, (\der_{v}\inn\ \Omega) $}     .
\eeq
Now, one can easily calculate the components of the
vector-valued form $\xt_{H\om}:=\xt^v{}_idx^i\otimes \der_v$
associated with $H\om$:
\beq
\mbox{$\tilde{X}$$^a_{\cdot k}$$ = \der^a_k H,$ \hspace*{1em}
$\tilde{X}$$^i_{a k}$$\delta^k_i = -\der_a H , $}
\eeq
and reveal that the natural parametrization of $\xt_{H\om}$:
\beq
\tilde{X}^v_{\cdot k} = \frac{\der z^v}{\der x^k} ,
\eeq
leads to the DW Hamiltonian field equations.
It means that $\xt_{H\om}$ also may be thought of as the
analogue of the canonical Hamiltonian vector field and
$H\om$ as the analogue of the canonical Hamilton's function.

Now, formally generalizing (25) one defines
the (tilded) bracket with $H\om$ to be (cf. next section)
\beq
\pbr{H\om}{\ff{p}}\tilde{}:=\xt_{H\om}\inn \dv \ff{p}.
\eeq
Calculating this bracket on extremals, that is using (1),
one obtains for any $p$-form $\ff{p}$
\beq
\mbox{{\bm $d$}}\ff{p}=\pbr{H\om}{\ff{p}}\tilde{}+d^{hor}\ff{p} .
\eeq
This extends the bracket form of equations of motion to
arbitrary forms and, in particular, reproduces the DW
Hamiltonian field equations if suitable forms are
substituted into the bracket. The tilde sign with a
bracket means that it is calculated on extremals, so that
the primitive fields cannot be factored out and, as
a consequence, the graded antisymmetry
of the bracket may be lacking.
This is because the successive action of a multivector
associated with $\ff{p}$ and a vector-valued form
associated with $H\om$ has, in general,
both graded commuting and graded anticommuting parts.

\section{Non-Hamiltonian forms and graded
derivation operators on differential forms}

The analytical restriction imposed on Hamiltonian forms
(cf. sect. (4))
excludes from the scheme constructed above some forms which
are interesting from the point of view of the canonical
theory under consideration.
For example, the DW Hamiltonian $n$-form in the theory of
interacting scalar fields $y^a$ in Minkowski space-time,
which is given by the Lagrangean density
$L= \frac{1}{2}\der_i y^a \der^i y_a - V(y^a), $
may be written as
$H\om = -\half *p^a\we p_a + V(y)\om $
in terms of the $(n-1)$-form momenta variables
$p_a=p_a^i\der_i \inn \vol$ which are Hamiltonian and
their Hodge duals $*p_a:=-p_a^i dx_i$ which are not.
In passing to
a quantum theory one will need to represent both
$p_a$ and $*p_a$ as certain operators on a Hilbert space and,
therefore, one should know how to calculate the Poisson brackets
with $*p_a$, which is impossible within the framework developed
so far. To do this would require associating,
through some extension of the map (13) given by
the polysymplectic form,
certain objects, more general than  multivector fields, to the
non-Hamiltonian forms, like $*p_a$ are.
Such a generalization is an objective of this section.
It is inspired by a recent work of
Vinogradov \cite{Vinogr} and further details
will be published elsewhere.

Most generally, one can associate to a $p$-form $\ff{p}$ the
vertical graded derivation operator of degree $-(n-p)$
\beq
\dd{n-p}_F\inn\ \Omega = \dv \ff{p}
\eeq
The term vertical (graded) derivation means
that $\dd{q}$ lowers the
vertical degree of a form on which it acts through the
generalized inner product $\inn$
(the sign which one omits in what follows for short)
by one and the horizontal
degree by $(q-1)$. Such a graded derivation operator
may  in principle be represented
by means of the  vertical $(q+p)$-vector-valued horizontal $p$-forms
$\xt \in \Lambda^{q+p}_p$ acting on forms through the generalized
Fr\"olicher-Nijenhuis inner product (cf. \cite{Vinogr}).
Given the form $\ff{p}$ there
always exists some graded derivation of degree $(n-p)$ associated with it in
the sense of (37),
which can be properly represented in terms of the suitable
multivector-valued form. Evidently, the graded derivation
operator associated with a given form is defined modulo addition of
primitive graded derivations $\dd{p}_0$ $(p=0,...,n)$ annihilating
the polysymplectic form:
\beq
\dd{p}_0 \Omega = 0,
\eeq
so that the symbol $\dd{n-p}$ in (37) actually denotes the equivalence
class of graded derivations of degree $(n-p)$ modulo primitive graded
derivations of the same degree.

\medskip


The simplest generalization of the formulae (25)
for the Poisson brackets enables us to define
the Poisson brackets of arbitrary forms as
follows:
\beqa
\pbr{\ff{p}_1}{\ff{q}_2}&:=& (-)^{(n-p)} \half
[\dd{n-p}_1  \dd{n-q}_2 +
(-)^{(n-q)(n-p)}\dd{n-q}_2 \dd{n-p}_1 ]\, \Omega \nn \\
& &       \\
&=&(-)^{(n-p)} \half [\dd{n-p}_1 \, \dv \ff{q}_2
+ (-)^{(n-q)(n-p)}
\dd{n-q}_2 \, \dv \ff{p}_1]  \nn
\eeqa
This definition ensures the same graded antisymmetry of the
bracket as in (27) and it reduces to (27) when both of
the graded derivation operators are representable
in terms of  vertical multivectors.
By a straightforward calculation one can also prove that
the Poisson bracket above possesses all the properties
of the Poisson bracket of Hamiltonian forms, eqs. (26)--(29).
Therefore, the space of horizontal forms is a Gerstenhaber
algebra with respect to the exterior product and the
Poisson bracket defined in (39).

As an example, we shall calculate the graded derivation of
degree $(n-1)$ associated with the non-Hamiltonian one-form
$*p_a=-p_a^i dx_i$. We may represent this graded derivation
(on the subspace of forms having a horizontal degree $(n-1)$)
by a vertical $n$-vector-valued one-form
\beq
\xt_{*p_a} :=   \frac{1}{(n-1)!}\xt_{*p_a}{}^v{}\uind{i}{n-1}{}_k\,
dx^k \otimes \der_v{}\lind{i}{n-1} .
\eeq
acting on forms through the generalized
Fr\"olicher-Nijenhuis inner product $\inn$ (its definition in
this case is apparent from eq. (42) below). It is given by
\beq
\xt_{*p_a}\inn \Omega = \dv *p_a ,
\eeq
or in components
\beqa
\xt_{*p_a}\inn\ \Omega &=&
\frac{1}{(n-1)!}\xt_{*p_a}{}^v{}\uind{i}{n-1}{}_k\,
dx^k \we \der_v{}\lind{i}{n-1}\inn\ (dp_a^i\we dy^a \we \om_i)  \\
&=& \nn n(-)^{c_{n-2}}\xt_{*p_a}{}^v{}\uind{i}{n-1}{}_k\,
\eps_i{}\lind{i}{n-1} dx^k \we \der_v \inn\ (dp_a^i\we dy^a)  \nn  \\
&=&
- dp^i_a\we dx_i \nn
\eeqa
where $c_p:=1+2+...+p$.
Therefore,
\beq
n(-)^{c_{n-2}}\xt_{*p_a}{}^v{}\uind{i}{n-1}{}_k\,
\eps_i{}\lind{i}{n-1} =
- \delta_a^v g_{ik}  .
\eeq

Let us calculate now  the bracket of $*p_a$ with an
arbitrary $(n-1)$-form $F=F^i\om_i$:
\beq
\pbr{F}{*p_a}=-\half[\, D_F \inn \dv *p_a
+ (-)^{n-1} \xt_{*p_a} \inn \dv F \,]
\eeq
where the vertical graded derivation of degree $-1$
associated with $F$, $D_F$, is represented by a vertical
bivector-valued one-form $\xt_F$,
\beq
\xt_F:=\xt^{vi}{}_kdx^k\otimes(\der_v\we \der_i)  ,
\eeq
acting on forms through the Fr\"olicher-Nijenhuis-type
inner product. From
$\xt_F\inn\ \Omega=\dv\ff{ }$
written in components it follows
\beqa
-\xt^{ik}_{a\cdot k}+\xt^{ki}_{a\cdot k}&=&\half
\der_aF^i,  \\
\xt^{ak}_{\cdot \; \cdot k}\delta^i_j-\xt^{ai}_{\cdot \; \cdot j}&=&
\half
\der^a_jF^i .
\eeqa
For calculating the bracket in (44)
one needs the components  $\xt_a^i{}_j$ along the momenta
directions only. Modulo addition of a primitive field
$\xt_0: \xt_0 \inn \Omega =0$,
the solution of (46) reads
\beq
\xt_a^{ij}{}_k = - \half \frac{1}{n-1}\delta^j_k
\der_a F^i \quad mod\, [\xt_0]
\eeq
Now, note that the components of $\xt_F$
representing  $D_F$  are, essentially,
given by the map (37)
only on the subspace of forms of horizontal
degree $(n-1)$.
On the subspace of forms of horizontal degree $p$
the operator
$D_F$ is rather represented by $\frac{n-1}{p} \xt_F$.
This follows from the observation that the operator
\beq
\bI_p := \frac{1}{p} (\delta^i_jdx^j\otimes \der_i)\inn,
\qquad \bI_0 := 1
\eeq
is the unit operator on the subspace of $p$-forms and
also from the requirement that when the $(n-1)$-form $F$
is Hamiltonian the graded derivation $D_F$ should be
equivalent to the derivation operator of degree $-1$ given
by the vertical vector field $X_F$ which is associated with
a Hamiltonian $F$ (cf. sect. (4)).

Taking these remarks into account, one obtains from (44)
\beq
\pbr{F}{*p_a}
= \der_aF^i dx_i .
\eeq
It is interesting to reveal the following property
\beq
\pbr{*p_a}{F}=*\der_a F^i \om_i = * \pbr{p_a}{F}
\eeq
which may be useful for representing the quantum
operator corresponding to $*p_a$.

Note, that the simplicity of the example choosen
above hides some tricky aspects of representation
of the equivalence classes of graded derivations
associated with forms in terms of the
multivector-valued forms. They are to be treated
elsewhere.

\medskip

\medskip

\centerline
{\sf It is a great pleasure for me to contribute to the volume in
honour of Professor D.D. Ivanenko.}
\centerline
{\sf This is his group of general
relativity  at Moscow University}
\centerline
{\sf where I started my way in theoretical physics ten
years ago.}


\begin{thebibliography}{99}

\newcommand{\bib}[1]{\bibitem{#1}}


\bib{Rund} H. Rund, {\em The Hamilton-Jacobi Theory in the Calculus of
Variations}, (D. van Nostrand Co. Ltd., Toronto, etc. 1966)

\bib{Kastrup83} H. Kastrup, Phys. Rep. 101 (1983) 1

\bib{Born34} M. Born, Proc. Roy. Soc. A143 (1934) 410
\bib{Weyl34} H. Weyl, Phys, Rev. 46 (1934) 505


\bib{Herm lie} R. Hermann, {\em Lie Algebras and Quantum Mechanics} (W.A.
Benjamin, inc., N.Y. 1970)

\bib{Garcia} P.L. Garc\'{i}a and A. P\'{e}rez-Rend\'{o}n, Comm. Math. Phys.
13 (1969) 24; Arch. Rat. Mech. Anal. 43 (1971) 101

\bib{Sniatycki} \'{S}niatycki,
Proc. Camb. Phil. Soc.  68 (1970) 475

\bib{Gold+Stern} H. Goldschmidt and S. Sternberg,
Ann. Inst. Fourier (Grenoble)
23, fasc. 1 (1973) 203; see also V. Guillemin and S. Sternberg, {\em Geometric
Asymptotics}, Math. Surv. 14 (AMS, Providence 1977) ch.4 $\S$8

\bib{Kij ea} J. Kijowski, Comm. Math. Phys. 30 (1973) 99; J. Kijowski and
W. Szczyrba, ibid. 46 (1976) 183; see also
J. Kijowski and W.M. Tulczyjew, {\em A Symplectic Framework for
Field Theories} (Springer-Verlag, Berlin etc. 1979)

\bib{Gaw} K. Gawedzki, Rep. Math, Phys. 3 (1972) 307; K. Gawedzki and
W. Kondracki, Rep. Math. Phys. 6 (1974) 465

\bib{Dedecker} P. Dedecker, in {\em Geometrie Differentielle},
Colloq. Internat. du CNRS, LII, Paris, CNRS 1953, p.17;
see also in {\em Lect. Notes. Math. 570} (Springer-Verlag,
Berlin etc. 1977) p.375 


\bib{Gotay ext} M.J. Gotay, in {\em G\'{e}om\'{e}trie Symplectique} $\&$ {\em
Physique Math\'{e}matique}, eds. P. Donato, C. Duval, e.a. (Birkh\"{a}user,
Boston 1991) p.160

\bib{Gotay multi1} M.J. Gotay, in {\em Mechanics. Analysis and Geometry:
200 Years after Lagrange}, ed. M. Francaviglia (North Holland,
Amsterdam, 1991) p. 203

\bib{Gotay multi2} M.J. Gotay, Diff. Geom. and its Appl. 1 (1991) 375

\bib{Gimmsy} M. J. Gotay, J. Isenberg, J. E. Marsden, R.
Montgomery, J. \'{S}niatycki and Ph. B. Yasskin: {\em Momentum maps and
classical relativistic fields: The Lagrangian and Hamiltonian
structure of classical field theories with constraints},
Preprint version, 1992.

\bib{Crampin} J.F. Cari\~nena, M. Crampin, L.A. Ibort, Diff. Geom. and
its Appl. 1 (1991) 345

\bib{Guenther87a} C. G\"{u}nther, J. Diff. Geom. 25 (1987) 23;
C. G\"{u}nther, in {\em Differential Geometric
Methods
in Theoretical Physics, Proc. XV Int. Conf.}, eds. H.D. Doebner and
J.D. Hennig (World Scientific, Singapore 1987) p. 14

\bib{Sardan} G. Sardanashvily, {\em Gauge Theory in Jet
Manifolds}, Monographs in Applied Mathematics, Hadronic Press, Inc,
Palm Harbor, 1993; see also
G. Sardanashvily and O. Zakharov, Int. J. Theor. Phys.
31 (1992) 1477; G. Sardanashvily
{\em Multimomentum Hamiltonian formalism
in field theory} preprint hep-th/9403172;
{\em Multimomentum Hamiltonian formalism in Quantum Field Theory}
preprint hep-th/9404001


\bib{Kan} I.V. Kanatchikov, {\em On the Canonical Structure of
the De Doder-Weyl Covariant Hamiltonian Formulation of Field
Theory I. Graded Poisson brackets and equations of motion},
Aachen preprint PITHA 93/41 (November 1993) and hep-th/9312162.

\bib{Kan in prep} I.V. Kanatchikov, in preparation;
see also {\em On the finite dimensional covariant Hamiltonian
formalism in field theory}, to be published in
{\em  New Frontiers in Gravitation}, ed. R. Santilli and
G. Sardanashvily, Hadronic Press, Palm Harbor, 1995





\bib{Arnold} V.I. Arnold, {\em Mathematical Methods of Classical Mechanics},
(Springer, N.Y. 1978);
see also b) G. Marmo. E.J. Saletan, A. Simoni,
B. Vitale, {\em Dynamical Systems. A Differential Geometric Approach to
Symmetry and Reduction} (John Wiley $\&$ Sons, New York etc., 1985)
and c) P. Libermann and Ch. Marle, {\em Symplectic Geometry
and Analytical Mechanics} (Reidel, Dordrecht, 1987)

\bib{Enriquez} A. Echeverr{\'\i}a Enr{\'\i}quez, M.C. Mu\`noz Lecanda,
Ann. Inst. Henri Poincar\'e 56 (1992) 27;
see also
b) {\em Geometry of Lagrangian First-order Classical Field Theories},
preprint, Barcelona July 1994.


\bib{SN} J.A. Schouten, Proc. Kon. Ned. Ak. Wet. (Amsterdam) 43 (1940) 449;
b) A. Nijenhuis, ibid. A58 (1955) 390, 403

\bib{Gerstenhaber} M. Gerstenhaber, Ann. Math. 78 (1963) 267; 
M. Gerstenhaber and S.D. Schack, in {\em Deformation Theory of Algebras and
Structures and Applications}, eds, M. Hazewinkel and M. Gerstenhaber (Kluwer
Academic Publ., Dordrecht 1988) p.11

\bib{FN} A. Fr\"{o}licher and A. Nijenhuis, Proc. Kon. Ned. Ak. Wet.
(Amsterdam) A59 (1956) 338, 350;
see also
F. Mimura, T. Sakurai and T. N\^{o}no, Tensor, N.S. 51 (1992) 193


\bib{Kolar} I. Kola\v{r}, P.W. Michor, J. Slov\'ak,
{\em Natural operations in differential geometry}, (Springer-Verlag, Berlin
etc. 1993)

\bib{Vinogr} A.M. Vinogradov, Matem. Zametki (Math. Notices) 47 (1990) 138;
A. Gabras, A.M. Vinogradov, J. Geom. Phys. 9 (1992) 75.


\end{thebibliography}
\end{document}